\documentclass[runningheads]{svmult}
\usepackage{makeidx}
\usepackage{graphicx}
\usepackage{subeqnar}
\usepackage{multicol}
\usepackage{physprbb}
\makeindex
\begin{document}
\title*{Current Uncertainties in the Use of Cepheids as
Distance Indicators}

\toctitle{Current Uncertainties in the Use of Cepheids as
Distance Indicators}

\titlerunning{Cepheids as Distance Indicators}

\author{Michael Feast}

\authorrunning{Michael Feast}

\institute{Astronomy Department, University of Cape Town, Rondebosch, 7701,
South Africa.\protect\newline
mwf@artemisia.ast.uct.ac.za}

\maketitle

\begin{abstract}
The methods of calibrating the luminosities of galactic Cepheids and 
determining Cepheid reddenings are considered
in some detail. Together with work on NGC4258
this suggests that the calibration presented is valid to about
0.1mag (s.e.) at least for Cepheids with near solar abundances. Metallicity
effects are considered, partly through the use of non-Cepheid moduli of
the LMC. To reduce the uncertainty substantially below $\sim 0.1 \rm mag$
will require extensive work on metallicity effects. 
Non-linearities in period-luminosity 
and period-colour
relations will also need to be considered
as will the need to distinguish unambiguously between fundamental
and overtone pulsators.

\end{abstract}

\section{Introduction}
The assigned title of this paper might suggest that Cepheids are poor or
untrustworthy distance indicators. In fact they are currently the best
fundamental distance indicators that we have. For (classical) Cepheids
within our own Galaxy, the zero-point of the period-luminosity relation in
the $V$-band (PL(V)) is known to $\sim 0.1$mag. However, if we wish to
confirm this level of accuracy and improve on it, we need to consider
a number of possible constraints and complexities.

I begin by considering the calibration of the zero-point of the PL(V)
relation within our own Galaxy and later discuss possible complications
with this relation, particularly those related to the chemical abundance of
Cepheids. In this connection the indirect calibration of Cepheid luminosities
through independent estimates of distances to other galaxies, particularly
the LMC, will be considered. 

In our own Galaxy there are basically
four methods that can be used to calibrate a PL relation; trigonometrical
parallaxes of Cepheids; statistical parallaxes of Cepheids; Cepheids
in clusters or binary systems, and; Pulsation parallaxes (Baade-Wesselink
type analyses). 

\section{Basic Relationships}
 In any determination of absolute magnitudes, the interstellar extinction
to the objects used must be taken into account. 
Reddenings of individual Cepheids
can be obtained from multicolour photometry (e.g. $BVI$ photometry [1][2]).
Relative reddenings of good individual accuracy can be obtained in
this way for Cepheids of a given metallicity, despite the method having
been questioned on non-quantitative grounds [120].
This is clear from a discussion of LMC and SMC data [2]. The spread
(standard deviation) in the derived 
individual reddenings, $E(B-V)$, after allowance
for small photometric uncertainties is only 0.03mag (LMC) and 0.02mag (SMC).
These are therefore upper limits to the intrinsic scatter in the method.
In addition the derived reddenings show no dependence on period [2]
\footnote{But note that the angle between the intrinsic amd reddening lines
becomes less favourable with decreasing intrinsic colour (i.e. decreasing
period). So the precision is a function of period.}.
Laney and Stobie [4] quote results showing good 
agreement between
individual BVI reddenings of galactic Cepheids and those determined in other
ways, although full details have not yet been published.
The zero-point of the BVI reddening system in our Galaxy is determined
from Cepheids in open clusters of known reddening. However as noted below
a knowledge of this zero-point is not necessary in some important
distance-scale applications.

Using BVI-based reddenings, or reddenings consistent with these,
period-colour relations (PC) can be constructed (e.g. [3][4]).
For the present discussion the following PC and PL relations have been adopted.
\begin{equation}
<B>_{o} - <V>_{o} = 0.416 \log P +0.314,
\end{equation}

\begin{equation}
<M_{V}> = -2.81 \log P + \rho
\end{equation}

The PC relation is for galactic Cepheids [4]. The slope of the 
PL relation is 
that for the LMC [5].
These are the basic relations used by Feast and Catchpole [6] in their work on
the Cepheid calibration using trigonometrical parallaxes. Later
we shall require a PC relation in $(V-I)$ and adopt [7][8] for galactic
Cepheids,
\begin{equation}
 <V>_{o} - <I>_{o} = 0.297 \log P + 0.427.
\end{equation}
Which is based on the same $BVI$ reddening system as equation (1). 

There is evidence in the literature of some misunderstanding regarding
the use of a PC relation. Both the PC and PL relations are approximations to
a period-luminosity-colour (PLC) relation
and both have significant scatter. In view of the scatter, using a PC 
relation does not produce the best possible estimates of the reddenings
of individual Cepheids. These can best be obtained from multicolour
photometry. However because of the relation between the PC and the PL
relations, through the PLC relation, deviations of PC-based reddenings
from true reddenings compensate for deviations of luminosities from the
mean PL relation. Thus the use of the two relations 
(equations (1) and (2)) together effectively
reduces the scatter in the PL relation. This reduction in 
scatter, i.e. in width of the PL relation, is by a factor
$((R/\beta) - 1)$, where $R$ is the ratio of total to selective
absorption ($A_{V}/E(B-V)$) and $\beta$ is the colour coefficient
in the PLC relation in $V$ and $(B-V)$. Since for the Cepheids, $R$ is 
$\sim 3.3$ and $\beta$ $\sim 2.5$, the PL width is, in this way, reduced by a 
factor of more than 3. This is important, not only in reducing the
scatter of estimates of the PL zero-point from individual stars, and hence
reducing the uncertainty in the mean value, but also in reducing the
effects of bias which are discussed below.

The reddening derived from a PC relation is a combination of the
true reddening with a measure of the deviation of the Cepheid from the
mean PC and PL relations. This being the case, negative PC reddenings can be 
expected and must be used, as must, of course, negative reddenings which are
simply due to statistical scatter. These negative reddenings are sometimes
dismissed in the literature as being ``unphysical", but this is due to a
misunderstanding of the PC/PL method.

The zero-point
of the PC relation is not of importance in distance determination
provided it is used consistently for both calibrating and programme
Cepheids (but note the exceptions discussed in sections 5 and 6).

\section{Trigonometrical Parallaxes}
 It is sometimes claimed or implied that since the trigonometrical
parallaxes 
which are currently available for many Cepheids have large percentage errors, 
they cannot be 
used to derive a trustworthy PL zero-point. This is not the case provided
there is a significant sample of stars and that
good estimates have been made of the standard errors of the individual
parallaxes. The massive and homogeneous astrometric survey carried out
 by the Hipparcos mission [9] produced data which appears to satisfy
these requirements. Nevertheless, the method of combining the data
has to be carefully chosen.

Consider a group of objects all of the same absolute 
and apparent magnitude and so
at the same (true) distance. 
The uncertainties in the absolute
magnitudes derived from their measured parallaxes ($\pi$) are proportional
to $\sigma_{\pi}/\pi$, where $\sigma_{\pi}$ is the standard error of $\pi$. 
Suppose $\sigma_{\pi}$ is the same for all the objects. 
Whilst the (weighted) mean parallax of the sample will be unbiased,
the absolute magnitudes from underestimated
parallaxes would have larger computed standard errors than those
from overestimated parallaxes and a weighted mean absolute magnitude
will be biased. This type of argument can be generalized as was done
by Lutz \& Kelker [10] and others (e.g.[11][12] see also [13]). The correction
to the derived mean absolute magnitude depends on the space distribution
of the objects concerned as well as on $\sigma_{\pi}/\pi$. In the case of
the Hipparcos parallaxes for Cepheids the necessary corrections would be
large for most of the stars. 
Since corrections of this type, unless very small, have considerable 
uncertainties, they are best avoided.
This can be
done by working in parallax space as will now be discussed.

If objects are selected by apparent brightness
(which will be so in the cases of interest here) there will be a selection 
bias if
there is a spread in absolute magnitude about the mean, or about a
relation such as the Cepheid PL relation. Bias of this kind was
discussed quantitatively by Eddington [14], Malmquist [15] and others.
The treatment in this section is from [16].

Consider first a group of objects with a mean absolute magnitude per unit volume
of $M_{o}$ and an intrinsic dispersion of $\sigma_{M_{o}}$. It is assumed there
has been no selection 
of the sample to be analysed
according to $\pi$ or $\sigma_{\pi}/\pi$. The method
of reduced parallaxes scales the measured parallaxes to the values they
would have at the same apparent magnitude. This can be written;
\begin{equation}
\overline{10^{0.2M}} = \sum{0.01 \pi 10^{0.2m_{o}} p}/ \sum{p}
\end{equation}
where the parallaxes are in milliarcsec, $m_{o}$ is the absorption free
absolute magnitude and p is the weight given by;
\begin{equation}
(0.01 \sigma_{T} 10^{0.2m_{o}})^{2} = 1/p
\end{equation}
and
\begin{equation}
\sigma_{T}^{2} = \sigma_{\pi}^{2} + b^{2}\pi_{M_{o}}^{2}(\sigma_{m_{o}}^{2}
 + \sigma_{M_{o}}^{2})
\end{equation}
In  equation (6), $\sigma_{T}$ is derived from the 
uncertainty in the parallax ($\sigma_{\pi}$), the intrinsic scatter in the 
absolute
magnitude ($\sigma_{M_{o}}$) and the uncertainty in the reddening corrected 
apparent magnitude ($\sigma_{m_{o}}$); also,\\
$b = 0.2 \log_{e} 10 = 0.4605$.\\
$\pi_{M_{o}}$ is the photometric parallax derived using the PL relation [12].
Put $x = (m_{o} - M_{o})$. Due to observational errors in $m_{o}$ and
intrinsic scatter in $M_{o}$, $x$ will differ from the true distance
modulus by $\epsilon$ (say). It is then evident that equation (4) yields
an estimate of;
\begin{equation}
\overline{10^{0.2M}} = \overline{10^{0.2(M_{o} + \epsilon)}} =
\overline{e^{bM_{o}}}.\overline{e^{b\epsilon}}
\end{equation}
Consider objects all of the same $m_{o}$ (and $x$).Then [15][17] ;
\begin{equation}
\overline{e^{b\epsilon}} = e^{0.5b^{2} \sigma_{t}^{2}} 
v(x - b \sigma_{t}^{2})/v(x),
\end{equation}
where,
\begin{equation}
\sigma_{t}^{2} = \sigma_{m_{o}}^{2} + \sigma_{M_{o}}^{2}
\end{equation}
and $v(x)$ is the frequency distribution of $x$ which would have been
observed if a complete survey had been made. It is important to note
that this is the case, whether or not the objects under consideration
actually form a complete survey. That is, the fraction of objects of
a given apparent magnitude, $m_{o}$, actually observed may be a function
of $m_{o}$ but this does not affect the quantity $\overline{e^{b\epsilon}}$.

Evidently at a given $m_{o}$ an unbiased estimate
of $10^{0.2M_{o}}$ is obtained by combining
equations (4), (7) and (8).  In general equation (8) is a
function of $x$ (that is $m_{o}$). Furthermore if we apply the method to 
large volumes of space (as is likely to be possible with GAIA parallaxes),
the function $v(x)$ may not be the same in all heliocentric
directions. If however we assume a constant underlying density distribution,
the r.h.s of equation (8) is independent of $x$ and becomes
$10^{-2.5b^{2}\sigma_{t}^{2}}$ (see e.g. [17] equation (9)). 
In this approximation,
the best unbiased estimate of $M_{o}$ is obtained by combining
equation (4) above, with;
\begin{equation}
M_{o} = 5 \log (\overline {10^{0.2M}}) + 1.151 \sigma_{t}^{2}
- 0.23 \sigma_{1}^{2}
\end{equation}
where $\sigma_{1}$ is the standard error of the derived value of $M_{o}$,
and the final term in equation (10) accounts for the
conversion between natural and logarithmic quantities.

The above discussion refers to a set of objects assumed to have a
mean absolute magnitude $M_{o}$ with a gaussian scatter. If 
this is not the case but instead the relative
absolute magnitudes of the objects are a function of some measured
quantity (e.g. in the case of Cepheids, the period) then
two cases need to be 
considered. If the measuring errors of this auxilliary quantity introduce
errors in $M_{o}$ which are small compared to $\sigma_{M_{o}}$, 
the formulation just given can, with obvious modification, be used to
find the absolute magnitude zero-point. If this is not the case a
different fomulation is required [16][17]. In the case of the
Cepheids the periods are 
usually known with good accuracy and their uncertainty
has a negligible effect on the predicted relative values of the
absolute magnitudes, so the formulation just given can be used.
 
Feast and Catchpole [6] analysed the Hipparcos parallaxes of Cepheids 
by the method of
reduced parallaxes. Similar results have been obtained by [18] [19].
Because they, [6], used the PC/PL approach discussed above to 
reduce the effective width of the PL relation, the bias term given in  
equation (10)
is very small (0.010mag). This would change their derived PL
zero-point ($\rho$) from $-1.43$ to $-1.42$. If, in an analysis of the
Cepheid data, the reddenings were derived in some other way then it would
be necessary to take into account the full (true) width of the PL(V) relation.
There is some evidence (e.g [3]) that Cepheids are distributed rather
uniformly through  a strip in the PL plane. If the half-width of this strip 
in magnitudes is $\Delta$, then for a constant space density distribution,
equation (8) becomes;
\begin{equation}
\overline{e^{b\epsilon}} = 3 sinh (2b\Delta)/2 sinh (3b\Delta)
\end{equation}
At longer periods in the LMC, $2\Delta$  
is approximately $0.7 \rm mag$ [3]. 
If this width applies to the calibrating Cepheids, the bias
correction terms amount to $\sim 0.05 \rm mag$. This much larger bias
shows the value of the PC/PL approach. Note that this bias remains the
same however accurate or numerous the individual parallaxes are. It is
perhaps worth noting that the need for
a bias correction of this type is not necessarily avoided by working
in magnitude space and applying a Lutz-Kelker type correction.

Whilst there are a large number ($\sim 220$) of Cepheids with Hipparcos 
parallaxes,
most of the weight in the reduced parallax analysis is in a relatively
small number of stars. The final result adopted [6] depends on the 26 stars 
of highest weight. This should now be corrected by the bias term in
equation (10) and results in the value shown in table 1 . If the overtone
pulsator, $\alpha$ UMi,
(see below) which carries about half the final weight
in this solution is omitted, a negligibly different result is obtained,
though of course with an increased standard error. 

An important recent
development has been the publication of a rather precise parallax
of $\delta$ Cephei itself
from HST observations [20]. The main uncertainty in this result probably
comes from the need to convert from relative to absolute parallax. 
Since this star was presumably chosen for measurement and analysis
because of its bright apparent magnitude and not (retrospectively) 
because of its parallax, a determination of its absolute magnitude
does not require a correction for Lutz-Kelker bias [16]. It will
however be subject to magnitude selection bias. A zero-point for the
PL relation from this one star is best derived using the PL/PC method
and equation (10). One then obtains the result shown in table 1.
The ``26" star solution [6] can now be improved by replacing
the Hipparcos parallax of $\delta$ Cep by a weighted mean of this value
with that from the HST result. This leads to the value also shown 
in the table 1. Incorporating
the result of Benedict et al. leads to a distinct lowering of
the uncertainty in the zero-point. The standard errors quoted are those
derived directly from the analyses. These, of course, have their own
uncertainties and 
Monte Carlo simulations by Pont [21] suggest that in the case
of the ``26" star solution of [6],  a more realistic estimate of the
standard error is 0.12  rather than 0.10 [8]. In view of this one might 
feel
that the uncertainty in the final value of table 1 (0.08) should
 be somewhat increased though it would seem unlikely to be greater than
0.10.
\begin{table}
\caption{Cepheid Trigonometrical Parallax Zero-Points (Bias Corrected)}
\begin{center}
\renewcommand{\arraystretch}{1.4}
\setlength\tabcolsep{5pt}
\begin{tabular}{ll}
\hline\noalign{\smallskip}
Method & $\rho$ \\
\noalign{\smallskip}
\hline
\noalign{\smallskip}
25 high weight & $-1.43 \pm 0.13$\\
$\alpha$ UMi fundamental & $-2.05 \pm 0.14$\\
$\alpha$ UMi 1st overtone & $-1.41 \pm 0.14$\\
$\alpha$ UMi 2nd overtone & $-0.97 \pm 0.14$\\
26 high weight & $-1.42 \pm 0.10$\\
$\delta$ Cep (HST) & $-1.32 \pm 0.10$\\
26 high weight revised & $-1.36 \pm 0.08$\\
\hline
\end{tabular}
\end{center}
\label{Tab1a}
\end{table}

Not all Cepheids pulsate in the fundamental mode and overtone pulsators
are most frequent amongst stars with short (fundamental) periods. Double
mode pulsators [22] provide the  period ratio of the fundamental
($P_{0}$) to the first overtone ($P_{1}$) for galactic Cepheids, e.g.
\begin{equation}
P_{1}/P_{0} = 0.720 - 0.027 \log P_{0},
\end{equation}
and this can be used to derive the 
fundamental periods of known overtone pulsators.
These overtone Cepheids may be identified using the Fourier components
of their light curves (e.g. [23]). They can also be identified in
an (observed) period - radius diagram using Baade-Wesselink type radii.
Early Baade-Wesselink work did not generally give radii of individual stars
of sufficient accuracy to do this. However more recent work (e.g.
using infrared photometry [24][25]) seems to be sufficiently consistent for 
this
purpose. It would therefore be important to obtain radii of high accuracy
for all the parallax stars of high weight. Only a few of them seem to have
the necessary data (e.g. $\beta$ Dor, $l$ Car, Y Oph and U Sgr are 
confirmed as fundamental pulsators in this way, and SZ Tau as an overtone
[24][25][26]). However the speculation [26] that the misidentification of
overtone Cepheids for fundamental pulsators amongst the high weight
parallax stars could have led to a significant overestimation of
Cepheid luminosities seems rather unlikely to be correct.

Polaris ($\alpha$ UMi) is treated in the analysis as a first overtone
pulsator on the basis of its derived absolute magnitude. If it were
either a fundamental or second overtone pulsator it would yield a PL
zero-point discrepant with the other high weight stars [6]. 
Evans et al. [27] have discussed other evidence that Polaris pulsates
in the first overtone, including the diameter of the star derived
using the interferometric angular diameter [28].
It is known from the LMC [29] that overtone Cepheids obey the normal
PLC relation (at their fundamental periods). However they are in the mean
brighter than the standard PL relation and intrinsically bluer
than a standard PC relation. Because of the PL/PC method of analysis
this means that the zero-point derived from overtone Cepheids
will be slightly too faint. In the present sample the effect of this
is expected to be negligible.

It has to be stressed that $\alpha$ UMi gives a PL zero-point 
in accord with that of the other high weight stars. The apparent 
discrepancy discussed by Di Benedetto [30] arises entirely because of
the PL zero-point he adopts ($-1.27 \pm 0.17$). However, this 
value is derived from a 
non-optimal selection of Cepheids (i.e. it does not contain all the high
weight Cepheids).
Note, however, that due to its larger error it is
is not significantly different from the values in table 1.

In later sections there will be a discussion of possible chemical abundance
effects on Cepheid luminosities. This is an area of some uncertainty.
The parallax Cepheids are all in the general
solar neighbourhood where the variation of chemical abundance amongst
young stars such as Cepheids is expected to be small. However,
abundance determinations for all the high weight Cepheids would be
desirable.

\section{Statistical Parallaxes}
The method of statistical parallaxes combines proper motions and 
radial velocities to obtain a PL or PLC zero-point. In common with the
method of reduced parallaxes discussed above, this method assumes that
the relative distances of the stars are known and only a scale value is
to be derived. In order to carry out an analysis of this type we require
a kinematic model. 
Both the proper motions [31] and the radial velocities [32] show clearly and
independently, the dominant effect of differential galactic rotation
on Cepheid motions in the Galaxy. Thus the required model must be based
on differential galactic rotation. This is even more apparent when one
considers that to a first approximation the amplitude of the differential
rotation effect in the proper motions is independent of distance whereas
for the radial velocities it is proportional to distance. Adjusting the
analyses for equality of the Oort constant ($A$) in proper motions and radial
velocities provides the best statistical parallax result for Cepheids. 
This is particularly the case since in this method the weight is
spread over a large volume of the Galaxy and so avoids problems due to
local deviations from an idealized model which almost certainly
occur. In this way zero-points were found [31][33] for a PLC and for a PL
relation. The zero-point of the latter, corrected for a possible
magnitude bias of $\sim 0.01$mag ( as discussed in section 3) is given in 
table 2.

In view of some discussions in the literature it is important to stress
that in deriving a PL zero-point from statistical parallaxes, there is a
great advantage,
as has just been mentioned, in treating the proper motions and the 
radial velocities
separately [31][32]. 

One can also attempt a solution using the solar motion obtained from a
combined discussion of solar motion and differential galactic rotation
using proper motions and radial velocities. In this way the solar motion
has a value which is averaged out over the whole large region of the
Galaxy covered by the 
proper motion (Hipparcos) and radial velocity Cepheids and is
not confined to a small region round the Sun where local deviations from
the idealized model may lead to false results. The resulting scale [34]
is only $0.04\pm 0.26$ mag larger than that just given. However the
standard error of this result is too large for the method to have any
significant weight.

The above discussion refers to the use of the systematic motions of the
Cepheids. In principle one can obtain a Cepheid scale from a comparison
of the dispersions about an adopted solution in radial velocities and
proper motions. However the velocity dispersion of Cepheids is small.
Thus any such solution will be sensitive to the treatment
of observational scatter in radial velocities and proper motions. It will 
probably also be sensitive to the effects of group motions. For these reasons
no attempt along these lines has been made here. A further discussion of
statistical parallax solutions is given in [8].

 The Cepheids used in the statistical parallax solutions cover a significant 
fraction of the galactic plane. Most of the stars lie in the range,
($R_{o} -3$) kpc to ($R_{o} +4$) kpc, where $R_{o}$ is the distance
of the Sun from the galactic centre. If there is a 
galacto-centric gradient in chemical
abundances of Cepheids over this range it might affect the PL and PC relations,
particularly the latter. Evidently the work now in progress on chemical
abundances of Cepheids (e.g. [35] etc.) should eventually allow us to take 
these effects
into account. However since the sample of Cepheids used in the statistical
parallax work is (roughly) centred on the Sun it may well be that any
effect in the final mean result will be small.

\section{Cepheids in Open Clusters}
 The (re)discovery of Cepheids in open clusters by Irwin [36] was
a major step in the Cepheid calibration problem. Whilst the use
of this method is of considerable importance, there are a number of
special problems associated with it. These are; (1) Uncertainty of
cluster membership; 
(2) Effects of reddening and photometric 
uncertainties; 
(3) Effects of metallicity;
(4) Absolute calibration of the cluster distance scale. 

(1). There are 30 open clusters or associations in our Galaxy which have
been listed as containing Cepheids [8]. Since that list was drawn up  
SZ Tau has been shown [37] from proper motions to be a non-member of
the cluster to which it was formerly assigned. 
In addition TW Nor is not used here because its cluster membership appears
doubtful [109].
Definite
confirmation of membership of several others would be very valuable.
It seems desirable that membership should be based on position
in the cluster, radial velocity and proper motion. In the past a
decision on membership has sometimes been made on the basis of whether
or not the derived Cepheid luminosity fitted with preconceived ideas.
This seems dangerously like an application of Merrill's [38] principle
which states that when the discordant observations are rejected the
remainder are found to agree very well. 

(2). The relative distances of the various clusters are obtained by
a main-sequence fitting procedure. Because of the steepness of the
main sequence this fitting procedure is very vulnerable to errors in 
the photometry or in the adopted reddenings. For instance an error
in $(B - V)_{o}$ of $\Delta (B-V)_{o}$ leads to an error in the
derived distance modulus of $\sim 6 \Delta (B-V)_{o}$
if the fitting is done in the $V,(B-V)$ plane. For some clusters,
distance moduli with standard errors as small as 0.02mag have been claimed. 
However,
even the adopted $(B-V)$ colours of photometric standard stars can vary
by 0.02mag or more between standard star observers [39]. Thus estimates of
the uncertainties of cluster moduli of $\sim 0.15$mag as
in Walker and Laney [40]
seem more realistic, and the errors could be greater in some cases.
If the cluster fitting is done  in the $V, (V-I)$ plane an error of
$\Delta (V-I)_{o}$ produces an error of $\sim 5 \Delta (V-I)_{o}$  in the
derived modulus.

In the case of the analysis of trigonometrical parallaxes and statistical
parallaxes of Cepheids it was pointed out that the zero-point of the
reddening system was not important so long as it was used consistently
for both the calibrating and programme stars. This is not the
case when calibrating Cepheids using clusters. Thus a change in the
reddening zero-point by $\Delta E$ changes the distance modulus
derived from $V, (B-V)$
by $\sim 6\Delta E$ and only $\sim 3\Delta E$ of this is recovered when
dereddening the Cepheid itself.

(3). The position of the main sequence is sensitive to metallicity effects.
A change in [Fe/H] of 0.1 dex leads to a change in absolute magnitude at
a given $(B-V)_{o}$ of $\sim 0.1$ mag,
e.g. [46]. It is generally
assumed that all the clusters 
containing Cepheids
are of solar metallicity or at least of
solar metallicity in the mean. The latter at least seems likely but it
has not been proved and further work on the  metallicities of the clusters
and their
Cepheids would be desirable.

(4). In the past the absolute calibration of the cluster distance scale
was based on an adopted distance modulus for the Pleiades. The value which
has generally been used, 5.57mag, is the value derived by van Leeuwen [41] 
by fitting
nearby field main-sequence stars with known parallaxes to the Pleiades
main sequence though this figure has been revised by others from time to time
[42][43]. It came as something of a surprise when the Hipparcos parallaxes of
Pleiades stars themselves gave a smaller distance modulus, 
$5.37 \pm 0.07$mag [44][45].
One reason for this surprise was that the van Leeuwen distance fitted 
rather well with theoretical
results for solar-metallicity main-sequences [46]. It has been suggested
that the Hipparcos distance can be reconciled with main-sequence theory
if the Pleiades are metal poor ([Fe/H] $= \sim -0.15$)[47]. There appears to 
be some 
evidence in Geneva-system photometry for such a suggestion [48] 
([Fe/H] $= -0.12 \pm 0.03$) but neither the
Stromgren photometry [49] ([Fe/H] $= +0.02 \pm 0.03$) nor
 spectroscopic abundances [50] ([Fe/H] $= -0.03 \pm 0.02$)
show evidence for significant metal poorness. 
These abundances are derived assuming that the Hyades cluster members have
[Fe/H] $= +0.13$.
Alternatively the Hipparcos mean
parallax of the Pleiades may have a greater uncertainty than given by its
formal error
or there is some problem with observations (see [8]) or theory. 
No final agreement on this point seems to have yet been reached.
However a rereduction of the Hipparcos data for the Pleiades stars
suggests a possible way out of this problem [121].

In view of this uncertainty it seems best at the present to base the cluster
scale on the Hyades for which there is an excellent Hipparcos-based parallax. 
The problem with this is that it
is generally agreed that the Hyades stars are slightly metal-rich, so a 
correction for this has to be made, if we make the common assumption
that the clusters with Cepheids are of solar metallicity in the mean.
The Hipparcos based distance modulus for the Hyades is
$(m-M)_{o} = 3.33 \pm 0.01$, [51] and the metallicity adopted by e.g. 
Pinsonneault et al. [46] is [Fe/H] = 0.13.  The 
theoretical metallicity correction adopted by these
latter authors then shows that the Hyades main sequence in $V, (B-V)$
corresponds to that expected for a solar metallicity cluster at
$(m-M)_{o} = 3.17$mag, or 3.12mag if the metallicity corrections of
Robichon et al. [45] are used. I adopt a mean value 3.14mag.
Since most work on clusters containing Cepheids is referred to
Turner's Pleiades main sequence [52], we need to see how this is affected by
the Hyades result.  The Pleiades - Hyades magnitude difference in a
$V,(B-V)$ diagram , corrected for reddening but not metallicity is 2.52mag
[53]. Thus the Turner main sequence is that expected for a solar metallicity
cluster at,\\
$(m-M)_{o} = 3.14 + 2.52 = 5.66.$\\
Adopting this value and assuming the clusters containing Cepheids 
which are listed
in [8] are in the mean of solar metallicity we obtain a PL
zero-point of,\\
$\rho_{1} = - 1.45 \pm 0.05$ (internal) mag.\\
Here SZ Tau 
and TW Nor have been omitted for the reasons given above. If the Hyades
metallicity suggested by Taylor [54] ([Fe/H] = +0.11) were adopted
we would obtain a brighter PL zero-point  (--1.47).
The error of this result is internal. The true error may well
be larger, partly due to uncertainties in the metallicity correction.
The standard deviation of the result is 0.26mag. 
Some of this is due to the width of the PL strip ($\sigma_{PL} = 0.21$ [5])
which comes in with full force here (unlike the case discussed in section 3).
Subtracting this quadratically gives 0.15mag as the standard deviation of the
cluster moduli. This agrees with a recent comparison of Baade-Wesselink
and cluster moduli by Turner and Burke [122] (their table 3) from which
one finds a
standard deviation of 0.14mag, presumably mainly due to the scatter
in the cluster moduli since the Baade-Wesselink results are believed
to have very high internal accuracy.

Whilst the adopted zero-point from clusters avoids the use of the
Pleiades modulus, the cluster method cannot be considered entirely trustworthy
until the problem of the Pleiades distance is fully understood.

Of the same nature as the cluster method is the use of physical
companions to Cepheids whose luminosity can be independently
estimated. This method has been used, notably by Evans and
collaborators [55][56]. At the present
time the accuracy obtained is not as good as that from other methods
(see [8]).

\section{Pulsation Parallaxes}
 An estimate of the luminosities of Cepheids can be made using 
pulsation parallaxes (Baade-Wesselink method). This method is dealt with
extensively elsewhere in this volume. The procedure normally used gives
results of high internal accuracy especially when implemented using
infrared photometry [24][25]
The method is currently being strengthened by interferometric measurements
of the angular diameters of Cepheids and their variation with phase 
[57][58][59][60].
It remains difficult to estimate in a realistic way  the true uncertainty
in the results from pulsation parallaxes which depend on possible systematic 
errors
in the derived radii and in the surface brightness estimates. For the present
discussion I have adopted a PL(V) zero-point derived from the results
of Laney [61] (see [8]).

\section{Summary of the Galactic Calibration}
The results discussed above are summarized in Table 2. The cluster and
pulsation parallax methods both have small internal errors but their
real (external) uncertainty is difficult to quantify. The results of
Monte Carlo simulations by Pont [21] suggest that in the case of the
``26" high weight 
parallax-solution Cepheids [6] the error might have been slightly
underestimated. That may still be the case here but it is unlikely to
be significantly greater than $\sim 0.1$mag. Both the trigonometrical and
statistical parallax methods seem rather robust. In the present paper an
unweighted mean has been adopted as best galactic zero-point and this is 
shown in Table 2.

\begin{table}
\caption{Galactic and NGC4258 Cepheid Zero-Points (Bias Corrected)}
\begin{center}
\renewcommand{\arraystretch}{1.4}
\setlength\tabcolsep{5pt}
\begin{tabular}{lll}
\hline\noalign{\smallskip}
No. & Method & $\rho$\\
\noalign{\smallskip}
\hline
\noalign{\smallskip}
1 & Trigonometrical Parallax & $-1.36 \pm 0.08$\\
2 & Statistical Parallax & $-1.46 \pm 0.13$\\
3 & Clusters & $-1.45 \pm 0.05$ (int)\\
4 & Baade-Wesselink & $-1.31 \pm 0.04$ (int)\\
  &     & \\
 & Unweighted Mean (1,2,3,4) & $-1.40$ \\
  &  & \\
5 & NGC4258 & $-1.17 \pm 0.13$ \\
  & & \\
 & Unweighted Mean (1,2,3,4,5) & $-1.35 \pm (0.05)$\\
\hline
\end{tabular}
\end{center}
\label{Tab1b}
\end{table}

\section{A Cepheid Zero-point from NGC4258}
A distance to the galaxy NGC4258 has been derived from the motions of
$\rm H_{2}O$  masers in the central region combined with a model [62].
In this way a distance modulus of $29.29 \pm 0.09$  was obtained. Newman
et al. [63] have obtained V,I data for Cepheids in this galaxy using the HST.
Reducing their data with the PC relation in $(V-I)$ given above and a 
PL(V) relation of slope $-2.81$ leads to a PL zero-point of
$-1.17 \pm 0.13$. Here the error in the maser distance has been
combined with the internal uncertainty in the Cepheid result.
In obtaining this value of the zero-point a small correction for
metallicity has been applied. HII region measurements [64] suggest that 
[O/H] is --0.05. A correction of 0.20 mag $\rm [O/H]^{-1}$
was adopted (see sections 9 and 10). Unless the adopted metallicity of the 
galaxy is
grossly in error or the metallicity effect much greater than assumed,
the metallicity correction is very small.

Including this zero-point
($-1.17\pm 0.13$) with the galactic
 values (Table 2) yields an unweighted mean of $-1.35 \pm 0.05$, which is
possibly the best current estimate of the zero-point for metal-normal
Cepheids. However it has been suggested recently [111] that 
model uncertainties lead to possible uncertainties in the mass of the central
black hole in NGC4258 of at least 25 percent and it remains to explore
what effect this has on the deduced distance.

\section{Metallicity Effects}
A remaining source of uncertainty in deriving distances of Cepheids is the 
effect of metallicity variations on the PLC, PL and PC relations, and 
on multi-band
intrinsic colours. In any distance derivation, the interstellar reddening
and absorption must be derived as well as a prediction of the
absolute magnitude of the star. The effect of metallicity change on PL relations
will vary with wavelength. The effect on the derived interstellar
absorption will depend on the method used for its derivation. For instance
there is good evidence from a comparison of the LMC and SMC that
Cepheids become bluer in $(B-V)$ at a given period with decreasing 
metallicity [43]. Thus a standard PC relation in $(B-V)$ will
give too small a reddening
for a metal-poor Cepheid. However if the reddening of a metal-poor
Cepheid is derived from a standard two-colour, $BVI$, plot, the reddening
will be too high (see e.g. [2]). If we knew precisely the dependence on
metallicity of all the quantities involved and had sufficient data we could
solve in any given case for the reddening and
metallicity of a Cepheid and
also for its luminosity and distance. An indication of how this this 
could be done in
practise using $BVI$ photometry was given in [65]. 

So far as the use of a PL relation to derive luminosities together with
either multi-colour data or a PC relation to derive reddenings is
concerned,
the metallicity problem may be broken down into three part.\\
1. A possible change in bolometric luminosity at a given period.\\
2. A possible change in colour at a given temperature.\\ 
3. A possible change of temperature at a given period.

Laney and Stobie [66] showed that at a given period the metal-poor Cepheids
in the Magellanic Clouds were slightly hotter than those in our Galaxy.
Laney[67] then showed that the radii of Magellanic Cepheids as determined
from Baade-Wesselink type analyses fitted the galactic period-radius
relation. These observations seem to show that at a given period the
bolometric luminosity of a Cepheid increases with decreasing metallicity.
However the effect is small and within the uncertainties of the observations.
Further work along these lines would be valuable. An effect on the
bolometric luminosity obviously affects the results at all wavelengths in
the same way. 

The effects of items 2 and 3 above, on reddening and luminosity
depend on the wavelengths and methods used. Here we consider the effects
when using $V,I$ photometry as in the HST work on 
extragalactic Cepheids. Other cases
have also been considered, e.g. [8]. 

The HST work essential uses a PC relation in $(V-I)$ and a PL(V) relation
to determine reddening and distance. There is some confusion in the
literature regarding metallicity effects using $V$ and $I$. This arises
because the effects of metallicity on equations (2) and (3) are such that
the changes affect the derived distance modulus in opposite directions.
It is thus important to consider these two equations together. A direct
test of this was made by Kennicutt et al. [68]. They observed Cepheids
in the galaxy M101 at different distances from the centre 
of the galaxy where abundances
had been estimated for HII regions. 
The abundance is above solar in the inner field and below solar
in the outer field.
Their results lead to a metallicity
effect on a distance modulus derived using equations (2) and (3) of
$0.24\pm 0.16 \rm mag \, [O/H]^{-1}$. This is in the sense that without the 
correction the distance of a metal-poor Cepheid would be overestimated.
This result suggests there is a small metallicity effect in the $V,I$ method.
However the uncertainty in the result is large.  It should
also be borne in mind that although there seems little doubt that
there is a strong metallicity gradient in M101, the absolute values
of the metallicities in the fields studied by Kennicutt et al. [68] 
remain somewhat uncertain (see their figure 2 and the accompanying
discussion).

Laney [67][69] (see Feast [8]) discussed Baade-Wesselink radii and colours of 
Cepheids in the Galaxy, the LMC and the SMC and these lead [8] to
an effect in the moduli of 
$\sim 0.09 \pm \sim 0.04 \rm (int)\, mag \,  [O/H]^{-1}$ in the same sense 
as the
Kennicut et al. correction. Much of the weight of the Laney result depends
on the SMC.

\section{Non-Cepheid Distances to the LMC}
  The LMC is of great importance for the Cepheid problem. It provides the
slope of the PL(V) relation currently in use. Also the LMC showed
clearly that when Cepheids are dereddened using three colour photometry,
the PL relation has a significant width which is reduced to within
the observational errors when a PLC relation is used [70]. The zero-point
of the LMC PL relation can be established if the LMC distance can be 
independently determined. However it is known that the LMC Cepheids
are metal-deficient compared with those in the solar neighbourhood, e.g. [71].
Thus a comparison of the Cepheid luminosities in our Galaxy and in the LMC 
can be an important test of metallicity effects on Cepheids. This is
obviously a major source of concern in the use of Cepheids as
general distance indicators. It is particularly important to note
that a non-Cepheid distance to the LMC does not give a PL
zero-point for normal metallicity Cepheids independent of some
knowledge of the metallicity effect.

The distance to the LMC is discussed elsewhere in this volume but it
is necessary to give here the basis for the present discussion
on the luminosities of LMC Cepheids.
In the following subsections some non-Cepheid methods of determining the
LMC distance will be considered. Whilst many of these methods appear
promising it should be remembered that none of them have yet been
subjected to the intense scrutiny that has been applied to the
Cepheids themselves. In each case, some of the issues that need
resolving before that particular distance indicator can be fully
relied on, are mentioned. Only
methods which are, or have been claimed to be, largely independent 
of theory are considered. For instance the magnitude of the 
Red-Giant-Branch tip seems to be a good indicator of relative
distances but requires an absolute calibration either from stellar evolution
models or through some other indicator (such as Cepheids).

\subsection{The RR Lyrae Variables}
RR Lyrae variables have long been regarded as valuable distance indicators.
However the dependence of their absolute magnitudes on metallicity has
been a matter for debate. Furthermore if globular clusters are taken as 
a guide [72] there is a significant spread in $M_{V}$ at a given metallicity.
Other papers in this volume discuss the RR Lyraes in detail and a full
discussion is not given here. The most important recent development
has been the publication by Benedict et al. [73] of a trigonometrical
parallax of RR Lyrae itself. This can be used together with equation (4) 
above, to obtain an estimate of the mean 
absolute magnitude of RR Lyrae stars of
this metallicity (Fe/H] = --1.39). Account needs to be taken of the fact
that at this metallicity globular cluster results suggest that RR Lyraes
fill a strip of width $\sim 0.4$mag. One obtains [16], $+ 0.64 \pm 0.11$
correcting for the resulting bias using equation (11) above.
Then,
adopting the relation;
\begin{equation}
M_{V} = 0.18[Fe/H] + \gamma
\end{equation}
from Carretta et al. [74] and a mean reddening-corrected apparent magnitude
of $V_{o}= 19.11$ at a metallicity of [Fe/H] = --1.5 for the LMC field
RR Lyraes, one obtains an LMC modulus of $18.49 \pm 0.11$.
Note however that this standard error should probably be increased,
possibly to $\sim 0.16$ due to the uncertainty in the bias correction
as applied to the one calibrating star.
Other RR Lyrae-based estimates of the LMC modulus are listed in [75]
where a mean of 18.54 was adopted. The uncertainty in this latter value
is probably somewhat over 0.10 mag. Whilst the determination of an
accurate trigonometrical parallax for RR Lyrae is a great step 
forward, the accuracy of
the LMC modulus derived from it must be limited if the spread in 
absolute magnitudes at a given metallicity is as great as that adopted above.
It may however be possible to use this parallax result together with an 
infrared period-luminosity relation [76] to obtain a more precise result if
this latter relation has a small scatter

\subsection{The Mira Variables}
Multi-epoch infrared photometry of Mira variables 
in the LMC shows that both carbon-rich
(C-type) and oxygen-rich (O-type) variables have a well-defined PL relation
in the $K$ band ($\rm 2.2\mu m$)[77]. For the O-Miras the relation can be 
written,
\begin{equation}
M_{K} = -3.47 \log P + \gamma.
\end{equation}
The scatter about this relation is only 0.13mag.
Miras in the SMC, in globular clusters, as well as those with spectroscopic
parallaxes from companions, all fit a PL(K) 
relation with the same slope [78][79][80].
The zero-point may be calibrated using Hipparcos parallaxes of Miras.
This yields, $\gamma = +0.86 \pm 0.14$mag [81] when 
small bias effects (see equation
(10), above) [16] are taken into account. A zero-point can also be obtained
from Miras in globular clusters. This method gives, $\gamma = 0.93 \pm 0.14$mag
[80]. To this may now be added a result from the parallaxes of OH-maser spots
in Miras obtained using VLBI 
[113]. The four Miras with distances from this method, together with 
infrared photometry [114] yield, $\gamma = +1.04$mag. The internal
standard error of this result is small (0.13mag) but the uncertainties
in the individual determinations suggest that this is an underestimate and
that the standard error of the mean is probably about 0.23mag. In
view of this, the last method is given half weight in combining the
three estimates. One then obtains $\gamma = +0.92$
and an LMC modulus of 18.56. If full weight
had been given to the third method the zero-point would only have been
increased by 0.02mag. The standard error of the adopted result is less than
0.10mag.

In globular clusters the periods of Mira variables are a function of
metallicity e.g. [82]. It is not clear whether, at a given period, the 
metallicity
of Miras differs from system to system. However there is some evidence that
the infrared colours 
of O-Miras at a given period are systematically different in the
LMC from those of Miras in the galactic bulge. This is likely to be due to 
weaker $\rm H_{2}O$ bands in the LMC stars [82][83]. This could be due either
to a deficiency of oxygen (an $\alpha$ element) or to a higher C/O ratio
resulting from an overabundance of carbon.
The effect of this on the PL(K) relation is not known empirically.
However theoretical work [110] suggests that, if anything, a general
metal deficiency, if not taken into account, will 
lead to an underestimation of the distance modulus.

It is perhaps worth pointing out that whilst Miras seem reliable distance
indicators in the case of the Magellanic Clouds, caution is require
if only a few such variables are identified in a system. In the LMC,
whilst the bolometric PL relation found at short periods is continued
out to 
periods of
$\sim 1000$days by dust-enshrouded Miras [84][85], there are a number of stars
with periods over 420 days which lie above this relation [77]
and some similar objects at shorter period. Whitelock [84][85] has
pointed out that, of these, 
those studied by Smith et al. [86] show evidence for surface
lithium and can be interpreted as hot bottom burning stars which
would not be expected to obey the PL relation. Note that the Mira-like
variable in IC1613 
which is clearly too bright for the PL relation [87]
has an unusually early spectral type for its period (641 days) and
is a likely candidate for a hot bottom burning object [85].

\subsection{Eclipsing Binaries}
Deriving distances from eclipsing binaries has much in common with the
determination of pulsation parallaxes by a Baade-Wesselink type analysis.
In both cases a stellar radius is combined with an estimate of the
surface brightness to obtain a luminosity. The method has been applied to
three LMC eclipsing variables [88][89][90] and rediscussion of some of these
results have been published [91][92]. In view of the spread in distance moduli
derived, Fitzpatrick et al. [90] suggest only that they lead to an LMC
modulus of $\sim 18.40$. Uncertainties and assumptions in the method used
have been discussed [91][75]. 

\subsection{SN 1987A}
Panagia [93] deduced a distance to the LMC centroid from  the ring round
SN1987A ($18.58 \pm 0.05$). 
The distance depends, amongst other things, on
the assumed ellipticity of the ring. A spectral-fitting expanding-atmosphere 
model givs a similar result though with considerable
uncertainty ($18.5 \pm 0.2$) [94].

\subsection{The Red Giant Clump}
The use of the red giant clump as a distance indicator has been much
discussed in recent years. As applied to the LMC this has led to conflicting 
results. Girardi and Salaris [95] investigated theoretically the
dependence of the clump absolute magnitude on age and metallicity. Coupling
their results with a population synthesis model of the LMC they obtained
an LMC modulus of 18.55. More recently Alves et al. [96] have applied the 
method 
in the K-band and find $18.49 \pm 0.04$. However the need
to assume theoretical age and metallicity corrections and to adopt an LMC
model, reduces the usefulness of the clump as a distance indicator [95].

\subsection{Open Clusters}
 It has long been realized that main-sequence fitting to young
clusters in the LMC provides a method to estimate its distance. Since
this is the same procedure as that used to derive distances to Cepheids
in open clusters in our Galaxy (section 5, above), the qualifications 
discussed there apply also in the case of LMC clusters.
A particularly interesting result is that derived from extensive work on
the LMC cluster NGC1866 by Walker et al. [97]. These authors deduce
a reddening-corrected distance modulus of $18.35 \pm 0.05$ for the cluster
and point out that if it lies in the plane of the LMC the mean modulus
of this galaxy is 18.33.  

The NGC1866 modulus is derived differentially with respect to the
Hyades, adopting the Hipparcos distance for this cluster and applying 
a correction for the metallicity effect. The comparison between the two
main sequences is made in the $V, (V-I)$ plane and a value of
$A_{V}/E_{(V-I)}$ of 2.08 was adopted. 
Walker et al. obtain reddening corrected
moduli of 18.37 and 18.33 for assumed values of [Fe/H] of $-0.30$ and
$-0.50$. Since these metallicities span the range of metallicities
measured in various ways for the cluster, they adopt an NGC1866 modulus of 
18.35. It is interesting to note (as can be deduced from the diagrams
in their paper) that the distance modulus of NGC1866 corrected for 
reddening but
not for the metallicity difference between it and the Hyades 
([Fe/H] = +0.13) is 
$\sim 18.9$. A comparison 
of this with the results for the two different assumptions
as to the metallicity of NGC1866 shows that the applied metallicity corrections
are highly nonlinear. These results may be compared with those
which are obtained using the (linear) metallicity corrections of
Pinsonneault et al. [46]. These are moduli of $\sim 18.5$ and $\sim 18.3$
for [Fe/H] of --0.30 and --0.50. Clearly the metallicity model is
crucial. 

One might be concerned that the relative abundances of the heavy elements
might be different in the Magellanic Clouds from the Sun.
Hill et al. [98] found no significant 
enhancement or depletion in the ratio of $\alpha$-elements to iron
for stars in NGC1866 ([$[\alpha/Fe] = 0.1 \pm 0.1$). 
However the situation is not entirely clear since 
depletion of the $\alpha$-element oxygen seems rather general in the
LMC [98][99][112].

Finally it is worth noting that the uncertainty assigned by Walker et al.
to their favoured modulus for NGC1866  (0.05mag) should probably be
regarded as an internal error only. 
The external error is likely to be larger due to the  effects of 
magnitude transformations and other causes. For instance they find from a
comparison of their HST data with overlapping ground based data that
there is a mean difference in colour of
$\Delta (V-I) = -0.07 \pm 0.06$(s.d)
in the sense, ground based minus HST data. They do not apply this as a 
correction to their HST data in view of the fact that
it is much reduced if outliers are omitted. However had they applied this
correction
their distance modulus would have been $\sim 0.35$mag greater for the
same adopted reddening.This follows since a comparison with the ground based
data [100] shows that the HST results are the relevant data set for
the main sequence fitting. This of course does not necessarily prove
that the cluster distance modulus is $18.35 + 0.35 = 18.70$. However it does
indicate the uncertainties encountered in this type of work.

\subsection{LMC Summary}
 Mean distance moduli from various types of objects are listed in
table 3. These can be combined in a variety of ways. This generally
leads to mean moduli near 18.5. In view of the various points discussed
above one should probably consider this to have a realistic standard error
of about 0.1 despite internal accuracies better than this being
claimed for some determinations. In the present connection we are
not concerned with the LMC modulus per se. We require a distance which can be 
compared with that derived from Cepheids so as to estimate the effects
of metallicity on the Cepheid scale. In doing this a remaining uncertainty
is whether or not the reddenings adopted in deriving the
moduli listed in table 3 are consistent with those used for the Cepheids.

\begin{table}
\caption{Non-Cepheid LMC Distance Moduli}
\begin{center}
\renewcommand{\arraystretch}{1.4}
\setlength\tabcolsep{5pt}
\begin{tabular}{llll}
\hline\noalign{\smallskip}
Object & Method & Modulus & Mean Modulus \\
\noalign{\smallskip}
\hline
\noalign{\smallskip}
RR Lyraes & Trig. Par. & $18.49 \pm 0.11$ & \\
          & Hor. Branch & $18.50 \pm 0.12$ & \\
          & Glob. Cl. & $18.64 \pm 0.12$ & \\
          & $\delta$ Sct &$18.62 \pm 0.10$ & \\
          & Stat. Par. & $18.32 \pm 0.13$ & \\
          &            & unweighted mean &18.51 \\
    & & &\\
Miras & & $18.56 \pm < 0.10$ & 18.56 \\
  & & & \\
Eclipsers & & $\sim 10.40$ & 18.40 \\
  & & & \\
Red Clump & V-band & 18.55 & \\
      & K-band & $18.49 \pm (0.04)$ & \\
    &    & unweighted mean & 18.52 \\
 & & &\\
SN1987A & & $18.58 \pm 0.05$ & 18.58 \\
 & & &\\
NGC1866 &  & $18.33 \pm (0.05)$ & 18.33 \\
  & & &\\
  & & & \\
 All  & & Unweighted Mean & $18.48 \pm (0.04)$\\
\hline
\end{tabular}
\end{center}
\label{Tab1c}
\end{table}

\section{Tests of Metallicity Effects}

Table 4 shows the adopted non-Cepheid LMC modulus. Also shown is
the Cepheid distance modulus of the LMC based on equations (2) and (3)
and with the zero-point of the PL(V) relation from Table 2 ($-1.35$)
without any metallicity correction,
with the corrections used by the HST Key project group [107],
$0.20 \rm mag \, [O/H]^{-1}$,
and that derived from the work of Laney.

It is clear that the various estimates are in better agreement than
we might reasonably have expected.

\begin{table}
\caption{LMC Moduli: Cepheid - Non-Cepheid Comparison}
\begin{center}
\renewcommand{\arraystretch}{1.4}
\setlength\tabcolsep{5pt}
\begin{tabular}{lll}
\hline\noalign{\smallskip}
Object & Method & Modulus \\
\noalign{\smallskip}
\hline
\noalign{\smallskip}
Non-Cepheid & & $18.48 \pm (0.04)$\\
& & \\
Cepheid & V.I Uncorrected & 18.60 \\
  &   V,I Laney Correction & 18.56 \\
  &  V,I HST Adopted Correction & 18.52 (Range 18.57 - 18.47)\\
\hline
\end{tabular}
\end{center}
\label{Tab1d}
\end{table}

Some caution has to be used with these results. For instance there is evidence
(see section 10.6)
that young objects in the LMC may be deficient in oxygen (an $\alpha$ 
element) relative to iron
and this could affect the luminosities of some calibrators.
There remains also the problem of the
depth of the LMC. It is generally assumed to be small, at least for young 
objects. But more evidence bearing on this is required, especially for
the old objects such as RR Lyrae stars which are used as LMC distance
indicators. It is worth recalling that the SMC Cepheids show evidence
of a considerable depth of this galaxy in the line-of-sight [3] and this
tends to preclude the use of the SMC for stringent tests of the Cepheid
scale and its metallicity dependence.

A similar comparison of Cepheid and non-Cepheid moduli to that described above
for the LMC can be made for other galaxies. Dolphin et al. [101] and
Udalski et al. [102] have published such discussions 
for IC1613 in which 
the Cepheids are believed to have a low metallicity ($ \rm [Fe/H] \sim -1.0$).
These discussions suggest a relatively small metallicity effect
using $V,I$
for Cepheids of metallicity lower than that of the LMC. Dolphin et al. [101] 
found
$-0.07 \pm 0.16 \rm mag\,[O/H]^{-1}$ and Udalski et al. [101] suggest there is
no significant metallicity effect.  In the case of IC1613 there is some 
uncertainty,
due amongst other things to the lack of good estimates of the metallicities
of the Cepheids and other objects used to derive distances.

The above discussion suggests that in $V,I$ there is a small but probably
significant metallicity effect, at least for Cepheids more metal rich than
the LMC. It is evident that further progress requires amongst other things
improved abundances for Cepheids in both our Galaxy and nearby galaxies.

In view of the uncertainty in the metallicity correction it is advisable
to avoid the need for using it if possible. This is the case for the galaxies
in the HST key project [107]. The mean metallicity of these galaxies, weighted
by their contribution to the finally adopted value of $H_{o}$ is close to solar
([O/H] = --0.08). It has been hypothesised by some workers that the metallicity
effect at $V,I$ could be as high as $0.6 \rm mag\, [O/H]^{-1}$. Whilst it
seems unlikely that it could be as high as this, even such a large value
will only have a small effect on a value of 
the HST key project $H_{o}$ if this is based on the
galactic (or NGC4258) calibration.

\section{Cepheids -- General Problems}

There is
evidence of a metallicity gradient for Cepheids 
in our own Galaxy e.g. [35] and 
this will need to be taken into account in future analyses of Cepheid
data (see e.g. sections 3 and 4 above). Abundance determinations are also 
important
since they can help distinguish  first- and later-crossing Cepheids.
Most Cepheids are expected to be second-crossing stars and abundance
analyses (see [103] and papers referenced there) which show that 
they have undergone first dredge-up, are
consistent with this. The Cepheid SV Vul does not seem to have undergone
first dredge-up [104] and is therefore likely to be a first-crossing 
Cepheid. Evidently chemical analyses together with accurate parallaxes
will be a powerful way of investigating the multiple crossings and their
effect on the use of Cepheids as distance indicators.

In most discussions of Cepheid luminosities it is assumed that the slope
of the PL(V) relation can be taken from the LMC. In using this in our
own and other galaxies, this assumes that the slope is independent of 
metallicity. The empirical evidence on this point is not strong.
Caldwell and Laney [5] found a slope of $-2.63 \pm 0.08$ for the SMC
Cepheids, The value (adopted in the present paper) for the LMC is
$-2.81 \pm 0.06$ [5]. Udalski et al. [105] found $-2.76 \pm (0.03)$ for the LMC.
The standard error is placed in brackets since much of the weight
of this determination
is in short-period Cepheids. Gieren et al. [106] obtained
$-3.04 \pm 0.14$ for Cepheids in the general solar vicinity using
pulsation parallax results. There is a slight hint of a trend
SMC, LMC, Galaxy i.e. a metallicity effect. The evidence for such a trend
is evidently marginal and requires confirmation. The slope is of importance
since, for instance, the weighted mean log-period of the Cepheids used
in the Hipparcos parallax solution is smaller than the weighted mean log-period
of the extragalactic Cepheids used to determine $H_{o}$, e.g. [107]. If the
Gieren et al. slope had been used this would have resulted in an
approximately seven percent increase in the 
parallax distance scale as applied
to the HST 
key-project galaxies. Adopting the OGLE slope [105] would have had only
a small effect.

In the present discussion it has been assumed that the reddenings of
Cepheids are derived from a PC relation. The relations in $B,V$ and $V,I$
(equations (1) and (3) above) are both based on the system of BVI reddenings
and should be compatible. However there still remains work to make
certain that these form a completely self-consistent set. 
Some estimates of the PC slope in $(V-I)$ are shown in Table 5.
The value for the LMC derived from Udalski et al. [105] is based
primarly on short period Cepheids and is uncomfortably different
from the galactic value.
It suggests the possibility of a change in slope at about 10 days [108].
This is currently a cause for concern. 
The LMC slope of Caldwell and Coulson [3] which is weighted to
longer periods that of Udalski et al. differs from the latter
and the LMC and SMC slopes also seem to differ. However none
of these differences
are vastly  larger than their standard errors.
The question of changes of slope with metallicity and at a period of 
about 10 days
thus remains to be finally settled.

\begin{table}

\caption{Slope of the PC$(V-I)$ Relation}
\begin{center}
\renewcommand{\arraystretch}{1.4}
\setlength\tabcolsep{5pt}
\begin{tabular}{ll}
\hline\noalign{\smallskip}
Method & Slope\\
\noalign{\smallskip}
\hline
\noalign{\smallskip}
Galaxy (Caldwell, Coulson) [7] & $0.297 \pm 0.014$\\
LMC (Udalski et al.) [105] & $0.202 \pm 0.037$\\
 &  \\
Difference & $0.095 \pm 0.040$\\
  & \\
LMC (Caldwell, Coulson) [3] & $0.318 \pm 0.054$\\
SMC (Caldwell, Coulson) [3] & $0.227 \pm 0.038$\\
  &  \\
Difference & $0.091 \pm 0.066$\\
\hline
\end{tabular}
\end{center}
\label{Tab1e}
\end{table}

I have chosen to adopt here
the galactic PC
slope in $(V-I)$
for two reasons. Firstly, as just mentioned, 
the determination of Udalski et al. [105]
is heavily weighted by short-period Cepheids. Such stars
have little weight in the zero-point point calibrations discussed earlier
or in the current applications of Cepheids in galactic
astronomy
and in the determination of $H_{o}$.
Secondly, it happens that the mean metallicity of the
galaxies studied in the HST key-project is close to solar when weighted
according to their contributions to the final value of $H_{o}$. 
It is thus of some interest to compare the present calibration
for metal-normal Cepheids with that adopted by the HST key-project
group [107].
Some years ago [115]
there was a difference of about 8 percent (0.17mag) between the
Cepheid scale derived from Hipparcos parallaxes [6] and that adopted
by the Key-project group. This difference has been essentially eliminated
by two factors. Firstly, the Cepheid zero-point for metal-normal Cepheids
adopted from the various estimates in Table 2 is 0.08mag fainter than
that derived from Hipparcos parallaxes alone by Feast and Catchpole [6].  
Secondly, though the Key-project group still adopt an LMC modulus of
18.50, they apply a metallicity correction which implies that their
scale for metal-normal Cepheids has been increased by 0.08mag. 
The true value of $H_{o}$, however,
still remains somewhat contentious e.g. [108].

\section{Very Short Period Cepheids}
  The discussions above have left out of consideration the very short
period Cepheids (periods less than $\sim 2$ days). For instance
the analysis of Udalski et al. [105], although heavily weighted to shorter
period stars, omits Cepheids with periods less than 2.5 days.  
However, very short period Cepheids are known to be 
numerous in the SMC and other young metal-poor systems and although
they are relatively faint, they are potentially important as
distance indicators for systems such as the metal-poor dwarf galaxies
[116].

In the SMC there is a steepening of the PL relations for periods
shorter than 2 days, as was originally pointed out by Bauer et al. [117].
The OGLE data 
for these very short period Cepheids
in the SMC [118]  has been used by Dolphin et al. [116] to 
derive 
slopes of --3.10 and --3.31 for the PL(V) and PL(I) relations of 
fundamental mode pulsators and --3.30 and --3.41 for the first overtone
pulsators (which constitute about half the OGLE sample).

Dolphin et al. [116] have also discussed the relative distances
of the SMC and the dwarf galaxies, IC1613, Leo A and Sex A.
The Cepheids in these systems are expected to be metal poor since the values
of $12 + \log (O/H)$ from HII regions are; SMC, 8.0; IC1613, 7.9;
Leo A, 7.3; Sex A, 7.6 [119]. From a comparison of the relative distances 
derived
from the very short period Cepheids in these systems with relative
distances from other indicators (RR Lyrae variables, RGB tip, red clump)
Dolphin et al. conclude that there is no significant effect of metallicity
on the luminosities of these very short period metal-poor Cepheids.

\section{Conclusions}
The present discussion suggests that the luminosities of metal-normal
Cepheids are now known within a standard error of 0.1mag or 
possibly less. However
it has to be recognized that deviations from the true value 
considerably in excess of one standard error are entirely possible
statistically. It is therefore very desirable to further strengthen
the empirical determinations. Some remaining issues that need to be
resolved are as follows.\\
1. Do the slopes of PL (and PC) relations at different wavelengths
 vary with metallicity?\\
2. Are there non-linearities in the PL and PC relations? Particularly is there
a significant slope difference between short and long period
($> \sim 10$ days)
Cepheids
that would seriously affect the calibration and use of PL and PC
relations?\\
3. Are reddening effects being correctly and consistently treated in
the calibration and use of Cepheids?\\
4. Is the reddening law being used really applicable to all Cepheids in our
own and other galaxies?\\
5. Can better empirical estimates be obtained of the effects of
metallicity variations on Cepheid luminosities and colours?\\
6. Do the relative abundances of heavy elements (e.g. the ratio of $\alpha$
elements, such as oxygen, to iron) affect Cepheid PL and PC
relations?\\
7. Are a significant number of 
calibrating or programme
Cepheids undiscovered overtone pulsators?\\
8. Can we distinguish (probably by spectroscopic analysis) first-crossing
Cepheids from others? Are the luminosities of such stars significantly
different from others of the same period? If so, are these stars
sufficiently numerous to create a problem for the use of Cepheids
as distance indicators?

\end{document}